\documentclass[reprint,amsmath,amssymb,aps,superscriptaddress]{revtex4-1}
\usepackage{graphicx}
\usepackage{dcolumn}
\usepackage{bm}
\usepackage[utf8]{inputenc}
\usepackage{amsmath, amsthm, amsfonts, amssymb}
\usepackage[svgnames]{xcolor}
\colorlet{color1}{NavyBlue}
\usepackage[colorlinks=true,allcolors=color1]{hyperref}
\usepackage{physics}
\usepackage{dsfont}
\usepackage{graphicx}
\usepackage{cleveref} 
\usepackage{xfrac}
\usepackage{mathrsfs}

\begin{document}

\title{Asymptotic horizon formation, spacetime stretching and causality}

\author{Carlos Barceló}
\email{carlos@iaa.es}
\affiliation{Instituto de Astrofísica de Andalucía (IAA-CSIC), Glorieta de la Astronomía, 18008 Granada, Spain}
\author{Valentin Boyanov}
\email{vboyanov@ucm.es}
\affiliation{Departamento de Física Teórica and IPARCOS, Universidad Complutense de Madrid, 28040 Madrid, Spain}
\author{Raúl Carballo-Rubio}
\email{raul.carballorubio@ucf.edu}
\affiliation{Florida Space Institute, 12354 Research Parkway, Partnership 1, Orlando, FL 32826-0650, USA}
\affiliation{IFPU - Institute for Fundamental Physics of the Universe, Via Beirut 2, 34014 Trieste, Italy}
\author{Luis J. Garay}
\email{luisj.garay@ucm.es}
\affiliation{Departamento de Física Teórica and IPARCOS, Universidad Complutense de Madrid, 28040 Madrid, Spain}
\affiliation{Instituto de Estructura de la Materia (IEM-CSIC), Serrano 121, 28006 Madrid, Spain}

\begin{abstract}
In this work we analyse asymptotically flat, spherically symmetric spacetimes in which an event horizon is present without any trapped surfaces. We identify two types of such spacetimes, each related to the asymptotic behaviour (in time) of one of the two degrees of freedom of the metric. We study the causal structure of both types, showing that one almost always has a Cauchy horizon beyond which it is extendable, while the other is inextendable but has two separate future null infinity regions on either side of the horizon. We also study what energy conditions can be satisfied by the matter around the horizon. Some of these spacetimes were first introduced in an earlier work in which semiclassical effects near black-hole horizons were analysed. Here we generalise this analysis to a larger family of geometries.
\end{abstract}
\maketitle
\section{Introduction}

Black hole spacetimes, a prediction of standard general relativity, are the most clear situation begging for an appropriate admixture of general relativity and quantum mechanics. The presence of trapped surfaces and their associated singularities make it necessary to seek out the quantum physics which is related to these geometric features. Black holes are therefore among the best testing grounds for theories attempting to mix these two ingredients. A very successful way to inquire about quantum effects in gravitational settings has been the use of quantum field theory in curved backgrounds. It was using this framework that Hawking arrived at his most famous discovery~\cite{Hawking1975}: black holes should emit thermal radiation and evaporate accordingly.\par
Quantum fields develop subtle effects when evolving through geometries which generate trapping horizons or through regimes in which the formation of trapping horizons is close to occurring. For instance, in a previous paper~\cite{BBCG19} we analysed and compared the form of the Renormalised Stress-Energy Tensor (RSET)~\cite{BD,Wald95} in several situations of this sort. Particularly, we confirmed the result discussed previously in~\cite{Barceloetal2008}, namely that when the generation of a horizon happens at a slow pace, e.g. when the collapse of a ball of matter occurs at low velocity, the RSET can acquire large values near the horizon. This implies that such hypothetical evolution could not be analysed in the framework of classical general relativity but would have to include semiclassical effects.\par
Another rather interesting situation analysed in~\cite{BBCG19} is one in which there is light-trapping behaviour without the presence of actual trapped surfaces. Since Hawking's original calculation, Hawking radiation appeared to be strongly tied up to the existence of a trapping horizon. However, in a series of papers~\cite{Barceloetal2006,BLSV06,Barceloetal2011a,Barceloetal2011b,BBCG19} it has been shown that from a purely geometrical point of view it is possible to have Hawking-like radiation without really generating any trapped surfaces. These works consider geometries which tend toward the formation of trapping horizons but only asymptotically in time or, more generally, producing by whatever means an exponential peeling of geodesics during a sufficiently long period of time~\cite{Barceloetal2011a,Barceloetal2011b}.

In this work we are interested in further exploring the properties of these particular geometries. Specifically, after presenting a set of geometries with the trapping behaviour mentioned above (Section~\ref{Sec:geometries}), we will start analyzing their distinctive causal behaviour (Section~\ref{Sec:causality}). The interesting feature of these spacetimes from a purely geometric perspective is that they contain no trapped surfaces yet form an event horizon, as we will discuss in detail. Initially we will present them as geometric ad hoc constructions, although later in the paper (in section ~\ref{Sec:energyconditions}) we will analyse in detail whether they could be obtained as solutions of Einstein equations for some plausible matter content, discussing the energy conditions that can be satisfied around the trapping region while supporting these configurations. Finally, we will also analyse the characteristics of the Hawking radiation that these configurations can generate.
 
\section{The geometries}
\label{Sec:geometries}

Let us start by writing a generic family of spherically symmetric metrics of the form
\begin{equation}\label{1}
ds^2=-f\mathop{dv^2}+2g\mathop{dv}\mathop{dr}+r^2\mathop{d\Omega^2},
\end{equation}
where $f$ and $g$ are generally functions of $v$ and $r$, or just of $r$ in static cases. We use an advanced null coordinate $v$ because we are interested only in the future causal structure of these spacetimes. We will also assume that the geometries are regular at $r=0$. We use this simplifying assumption to avoid causal aspects associated with singularities and concentrate just on those due to the presence of horizons. However, we note that our analysis can be generalised straightforwardly to geometries with singularities, so long as they do not overlap with the region of light-ray trapping we will study. We restrict our analysis to this local region, which can easily be inserted into a spacetime with a different global structure with much the same consequences.\par
We will work with two types of geometries, both of which trap outgoing light rays, but are otherwise quite different from one another. Let us present these two cases by first looking at two static configurations, which will later become the asymptotic limit in time of our dynamical models discussed in section~\ref{s3}.\par

\subsection{Static configurations}
\label{Sec:static}

The two cases which we will study originate from a simple consideration. From the line element \eqref{1} with $f$ and $g$ depending only on $r$, the equation which governs the paths of the outgoing light rays is
\begin{equation}\label{2}
\frac{dr}{dv}=\frac{1}{2}\frac{f(r)}{g(r)}.
\end{equation}
From this equation, it is apparent that these null trajectories do not distinguish between a situation in which $f$ is zero (as it occurs for some values of the radial coordinate in Schwarzschild, Reissner-Nordström, or similar spacetimes) and one in which $g$ diverges. Needless to say, the two situations are physically quite different in spite of this.\par
In these static configurations we will assume that in either case the right-hand side of \eqref{2} is zero at some radius $r_{\rm h}$ and that it can be expanded in a power series around this point approaching both from the inside \mbox{$r<r_{\rm h}$}, 
\begin{equation}\label{3}
\frac{1}{2}\frac{f(r)}{g(r)}= k_1(r_{\rm h}-r)+k_2(r_{\rm h}-r)^2+\cdots,
\end{equation}
and from the outside $r\geq r_{\rm h}$,
\begin{equation}\label{4}
\frac{1}{2}\frac{f(r)}{g(r)}= \tilde{k}_1(r-r_{\rm h})+\tilde{k}_2(r-r_{\rm h})^2+\cdots.
\end{equation}
Since we want the only zero of these expressions to be at $r_{\rm h}$ and we want to avoid creating a trapped region, we require that the first non-zero coefficients $k_i$ and $\tilde{k}_j$ of both series be positive. If $g(r)\simeq {\rm const.}$ around $r_{\rm h}$, then we have a black hole which allows ingoing causal trajectories across $r_{\rm h}$ but not outgoing ones. On the other hand, if $f(r)\simeq {\rm const.}$, then $g(r)$ diverges as the inverse of a polynomial, which results in the same behaviour for outgoing light rays as before (since $f/g$ is the same), but for ingoing ones there is a difference: they are actually unable to cross the surface $r=r_{\rm h}$ either. This can be deduced from the expressions which describe their paths in this coordinate system, namely the geodesic equations for their radial trajectory $(v(\sigma),r(\sigma))$,
\begin{equation}\label{5}
v={\rm const},\quad
\ddot{r}=-\frac{\partial_r g(r)}{g(r)}\dot{r}^2\simeq\frac{m}{r-r_{\rm h}}\dot{r}^2,
\end{equation}
where the dot denotes the derivative with respect to the affine parameter $\sigma$, and $m$ is the order of the first non-zero term in the expansion \eqref{4}. Integrating this equation allows one to see that the affine parameter reaches an infinite value when the ingoing ray gets to $r_{\rm h}$ (e.g. for $m=1$, $r-r_{\rm h}\propto e^{-c\sigma}$, with $c>0$), indicating that, as seen from the outside, this surface is actually an asymptotic region (i.e. part of future null infinity). The interior region $r\le r_{\rm h}$ is therefore entirely separate from the exterior spacetime.\par
\begin{figure}
	\centering
	\includegraphics[scale=.6]{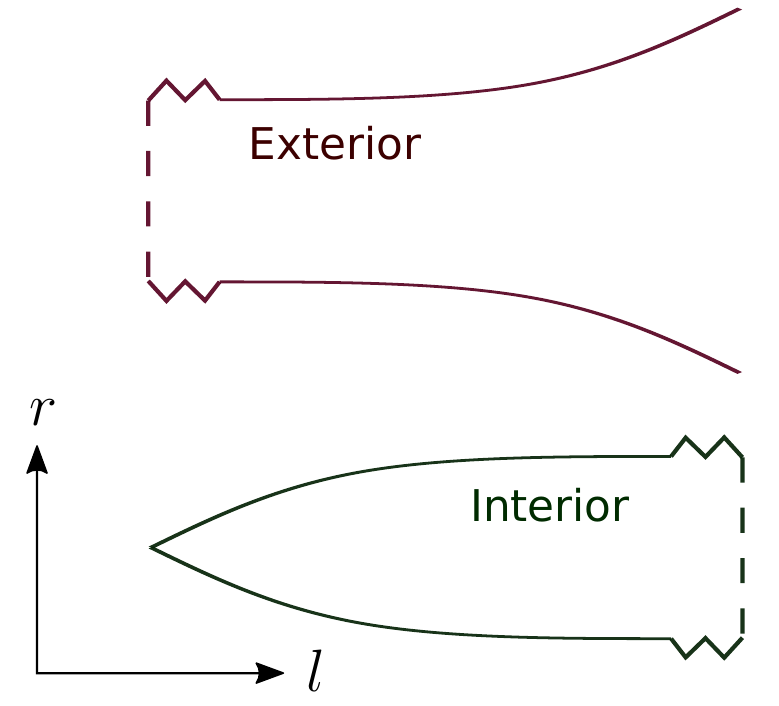}
	\caption{A qualitative representation of the relation between the radial coordinate $r$ and the proper length in the radial direction $l$ for a geometry in which $g$ diverges at some radius $r_{\rm h}$. This divergence corresponds to an infinite stretching of $l$ which completely severs the interior and exterior geometries.}
	\label{f4}
\end{figure}

To understand this situation better, we remind the reader that there is a more well-known spacetime in which $g$ diverges: that of a traversable wormhole. Particularly, this same configuration would be a standard spherical wormhole~\cite{VisserWH} if the expansion \eqref{4} had a leading term of order $(r-r_{\rm h})^{k}$, with $0<k<1$, as can be seen by calculating the proper radial length $l$ in slices of constant Schwarzschild time (defined by $dt=dv-(g/f)dr$) and expressing the radial coordinate $r$ as a function of $l$ around $r_{\rm h}$. On the other hand, when the leading order in the series is 1 or greater, as in our working case, the proper length diverges and space becomes infinitely stretched around the neck of the wormhole, becoming an infinite tube. Therefore, the static geometry we are considering here actually consists of two disconnected spacetimes, both having one infinite tubular ending (see Figure~\ref{f4}).

\subsection{Including time-dependence}\label{s3}

Having discussed these static spacetimes, we will now include a time-dependence in the metric functions $f$ and $g$ in order to push the formation of the apparent horizon/asymptotic region at $r_{\rm h}$ out to the limit $v\to\infty$. In a previous work \cite{BBCG19}, we used several such spacetimes (modeled after collapsing matter) in order to study the relation between characteristics of the geometry around the horizon and Hawking radiation. To do so, we analysed what particular types of time-dependence are necessary in order to trap outgoing light rays within a finite spatial region. Here, we will briefly summarise and expand upon these results, and analyse the causal structure of the resulting geometries.\par
\begin{figure}
	\centering
	\includegraphics[scale=.6]{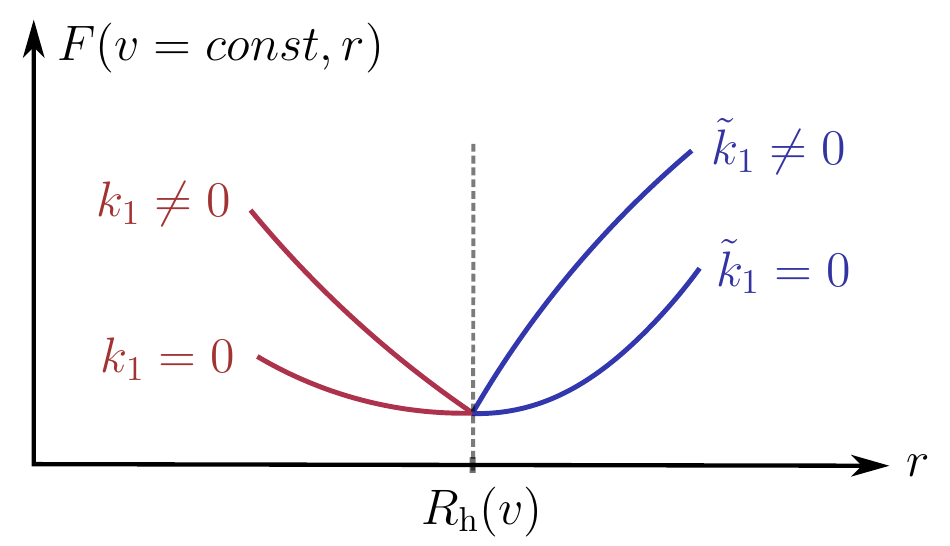}
	\caption{Slice at constant time ($v=const.$) of the generalised redshift function $F(v,r)$ around $R_{\rm h}(v)$. There is a discontinuity in the first derivative of this function at $R_{\rm h}(v)$ if either $k_1\neq 0$ or $\tilde{k}_1\neq 0$, which through the Einstein equations can translate into either a thin shell of matter, or into a discontinuity in pressure, depending on how the two individual degrees of freedom ($f$ and $g$) of the geometry which comprise $F(v,r)$ behave.}
	\label{f2}
\end{figure}

We will start with some definitions. First, we will call the right-hand side of eq. \eqref{2} the \textit{generalised redshift function} $F$,
\begin{equation}\label{6}
F(v,r)\equiv\frac{1}{2}\frac{f(v,r)}{g(v,r)}.
\end{equation}
We will assume that this function has a minimum in $r$ at a moving point $R_{\rm h}(v)$, and that it can be approximated by a series expansion on either side,
\begin{equation}\label{7}
F(v,r)=\delta(v)+k_1[R_{\rm h}(v)-r]+k_2[R_{\rm h}(v)-r]^2+\cdots
\end{equation}
for $r<R_{\rm h}(v)$, and
\begin{equation}\label{8}
F(v,r)=\delta(v)+\tilde{k}_1[r-R_{\rm h}(v)]+\tilde{k}_2[r-R_{\rm h}(v)]^2+\cdots
\end{equation}
for $r\geq R_{\rm h}(v)$ (see Fig. \ref{f2}). $\delta(v)$ is a function of $v$ which decreases and tends to zero in the limit $v\to\infty$. The function $R_{\rm h}(v)$ tends to a point $r_{\rm h}$ in the same limit. Our only simplifying assumption will be that the first non-zero coefficients $k_i$ and $\tilde{k}_j$ of the expansion on either side, aside from being positive, are approximately constant at large times (or, equivalently, that they tend to a constant at least as quickly as $R_{\rm h}(v)$).\par
It is worth mentioning that if either or both $k_1$ or $\tilde{k}_1$ are non-zero, then the function $F$ is continuous but not smooth at $r_{\rm h}$. Through equation \eqref{6} we see that this translates into a sharp peak in either $f$ or $g$ (or both) in slices of constant $v$. The Einstein tensor for the metric \eqref{1} only has a second partial derivative of $f$ with respect to $r$ (in its angular components), meaning a peak in $f$ corresponds to a spherical thin shell of matter. If the peak is in $g$, the tensor is only discontinuous, and so are the matter density, flux and stress seen by any observer. We note that the geometry is perfectly regular in spite of this discontinuity, unlike what one might expect in e.g. a static stellar configuration, where a jump in pressure leads to a singularity.\par
Regardless of the presence of the non-smooth peak (and its corresponding non-zero surface gravity), we will call all geometries in which $F$ has one zero (or an appropriate tendency to produce one zero; see the discussion below) and is positive everywhere else, \textit{extremal} configurations. We use this name because of a shared characteristic they have with the standard extremal black hole solutions: the presence of an outer and inner horizon which degenerate to the same radial position. We only extend the standard definition by allowing for a non-zero surface gravity on either side of the horizon.\par

If both $\delta(v)$ and $d_{\rm h}(v)=R_{\rm h}(v)-r_{\rm h}$ tend to zero sufficiently fast, then after some point in time some of the light rays which are inside the sphere of radius $r_{\rm h}$ remain trapped inside, the outermost of which defines the event horizon. In \cite{BBCG19} we assumed that these functions tend to zero at the same rate (e.g. $e^{-v}$, $1/v$, etc.), and the condition for light-ray confinement turned out to be a relation between this rate and the order of the first non-zero coefficient (and its value if the order is 1) in the expansion of $F$ for the interior \eqref{7}. Specifically, we showed:
\begin{itemize}
	\item If $\delta(v)\sim d_{\rm h}(v)\sim 1/v^n$, then light rays are trapped if the power $n$ and the order of the first non-zero coefficient in \eqref{7}, which we will call $m$, satisfy
	$$n-1>\frac{1}{m-1}.$$
	\item If $\delta(v)\sim d_{\rm h}(v)\sim e^{-\alpha v}$, then we can have any \mbox{$m\geq 1$}. If $m=1$, there is the additional condition \mbox{$\alpha>k_1$}.
	\item If these functions decay more quickly that an exponential, then there are no restrictions to the series~\eqref{7}.
\end{itemize}
These results were obtained by analysing the large $v$ limit of the solutions of eq. \eqref{2} for $r<R_{\rm h}$. With the series \eqref{7} we can write this equation as
\begin{equation}\label{9a}
\frac{dr}{dv}=\delta(v)+k_1[R_{\rm h}(v)-r]+k_2[R_{\rm h}(v)-r]^2+\cdots
\end{equation}
In this work we assume that the two functions $\delta(v)$ and $d_{\rm h}(v)=R_{\rm h}(v)-r_{\rm h}$ decay to zero independently, which leads to a generalisation of the above rules.\par
Let us first see the case in which $k_1\neq 0$. We write equation \eqref{9a} up to leading order as
\begin{equation}
\frac{dr}{dv}\simeq\delta(v)+k_1d_{\rm h}(v)-k_1(r-r_{\rm h}).
\end{equation}
From the functions $\delta(v)$ and $d_{\rm h}(v)$ on the right-hand side (rhs) we only need to consider the one which decays more slowly for the asymptotic solution. For example, if the slower of the two decays as $b\mathop{e}^{-\alpha v}$ (with $b$ and $\alpha$ some positive constants), then we can ignore the other one and obtain solutions of the form
\begin{equation}
r-r_{\rm h}\simeq -\frac{b}{\alpha-k_1}\mathop{e^{-\alpha v}}+c\mathop{e^{-k_1 v}},
\end{equation}
where $c$ is an integration constant. There are trapped solutions, which approach $r_{\rm h}$ asymptotically from below, only if $k_1<\alpha$: they correspond to the values $c<0$. On the other hand, if the slower of the two functions $\delta(v)$ and $d_{\rm h}(v)$ goes to zero more quickly than an exponential, then we again have a solution with a leading-order term $c\mathop{e^{-k_1 v}}$ and corrections which decay much faster, asymptotically recovering the same solutions as above for any value of $k_1$. Finally, if the slower of the two functions goes to zero more slowly than an exponential, e.g. as $1/v^n$, then there are no trapped solutions at all.\par
Now let us see the case in which $k_1,\dots,k_{m-1}=0$ and $k_m\neq0$. Equation \eqref{9a} becomes
\begin{equation}\label{12}
\frac{dr}{dv}\simeq\delta(v)+k_md_{\rm h}(v)^m+k_2(r_{\rm h}-r)^m+\cdots,
\end{equation}
where we have omitted the cross-terms in the leading order. If we assume the $(r_{\rm h}-r)^m$ term dominates the rhs, we obtain solutions of the type
\begin{equation}
r-r_{\rm h}\sim -\frac{1}{(v-c)^{\frac{1}{m-1}}},
\end{equation}
where $c$ is again an integration constant, and we have omitted a positive constant multiplying factor. These solutions are consistent with the assumption used above to obtain them as long as both $\delta(v)$ and $d_{\rm h}(v)^m$ decay at least as quickly as $1/v^n$, with
\begin{equation}\label{14-2}
n-1>\frac{1}{m-1}.
\end{equation}
On the other hand, if we assume one of the terms $\delta(v)$ or $d_{\rm h}(v)^m$ dominates the rhs of eq. \eqref{12}, then we get a solution of the type
\begin{equation}\label{15}
r-r_{\rm h}\sim\int^{v}\mathop{dv'}\max[\delta(v'),d_{\rm h}(v')^m],
\end{equation}
where the maximum is taken at sufficiently large $v$ to be in the asymptotic regime of the two functions. This solution is again only consistent with the assumption for the rhs of the differential equation \eqref{12} if the larger of the two functions decays at least as quickly as $1/v^n$, with $n$ satisfying \eqref{14-2}.\par
In summary, the generalisation of the rules in the above itemised list for this case is fairly simple: they are the same but must be satisfied by the two functions $\delta(v)$ and $d_{\rm h}(v)^m$ independently, or equivalently, by the one which goes to zero more slowly.\par
In these cases, the fact that the confined light rays do not reach the exterior future null infinity indicates the presence of an event horizon. This horizon's surface is described by the trajectory of the first trapped light ray, which can be seen to correspond to the solution \eqref{15}. Any outgoing rays which are outside it reach the asymptotically flat exterior region, and their dispersion is related to the presence and temperature of Hawking radiation. As for those on the inside, they must go to a different asymptotic region. How they end up depends on whether the asymptotic approach to zero in $F$ is due to a zero in $f$ or a divergence in $g$, as we will now see.\par

\section{Causal structure}\label{Sec:causality}

\subsection{Causal structure for finite $g$}

\begin{figure}
	\centering
	\includegraphics[scale=.55]{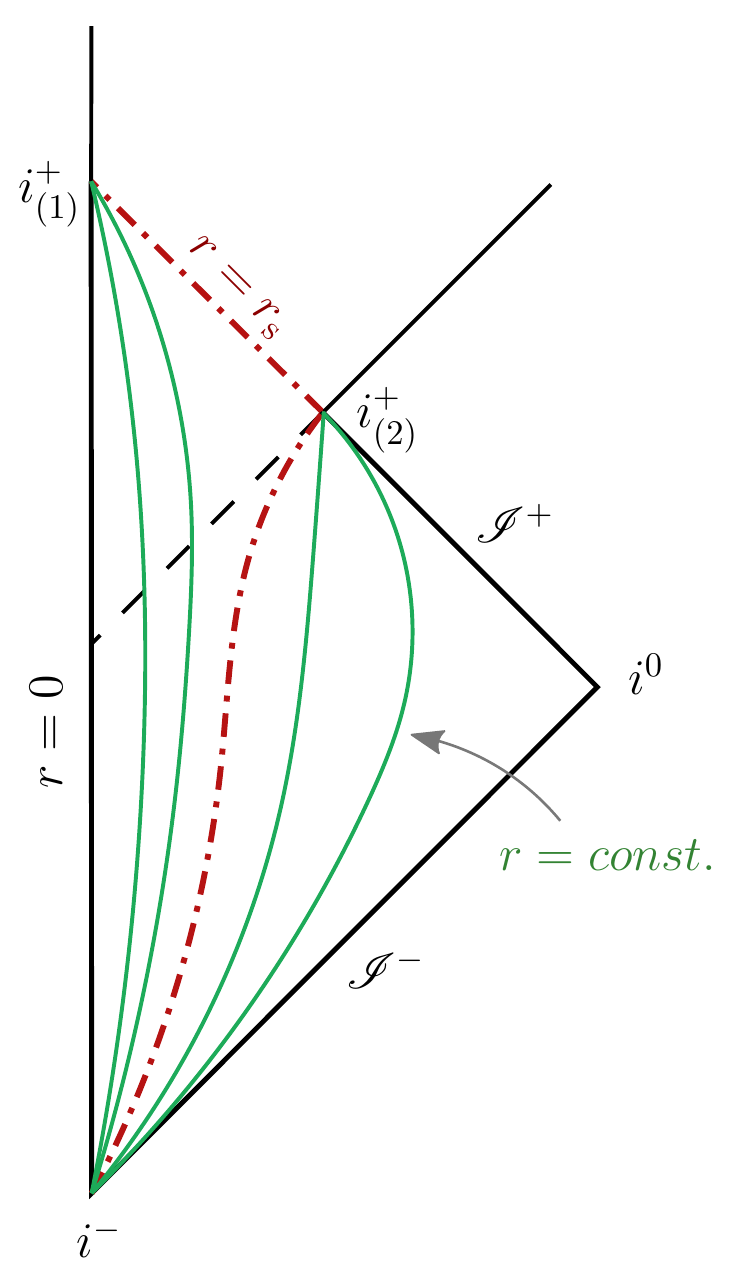}
	\caption{Conformal diagram of the spacetime with $g\simeq1$ and $f$ given by \eqref{7} satisfying the appropriate conditions for light-ray trapping. The dashed line is the event horizon, corresponding to the first trapped outgoing light ray. The dash-dotted line is the surface $r=r_{\rm h}$, which is described by a timelike curve and the Cauchy horizon. The curves to the left of $r_{\rm s}$ correspond to surfaces of $r={\rm const.}<r_{\rm h}$, while to the right they are $r={\rm const.}>r_{\rm h}$. The lines outside the conformal triangle indicate the need to extend the spacetime.}
	\label{f1}
\end{figure}

Let us assume that any light-ray trapping is due to an approach to zero in $f$ of the form \eqref{7}, and that $g$ remains finite (we will in fact assume $g=1$ for simplicity). The causal structure of the spacetime in this case almost always ends up being the same as that of an extremal (regular) black hole, shown in Fig. \ref{f1}. In the limit $v\to\infty$, the surface $r=r_{\rm h}$ becomes a Cauchy horizon, beyond which the geometry is extendable.\par

To show this, we can turn to one of the geodesic equations for a radial trajectory $(v(\sigma),r(\sigma))$ in our metric,
\begin{equation}\label{9}
\ddot{v}+\frac{\partial_rf}{2}\dot{v}^2=0.
\end{equation}
If the first non-zero coefficient in \eqref{7} is $k_1$, then the solution to this equation close to $R_{\rm h}(v)$ is
\begin{equation}\label{10}
k_1\dot{v}_0(\sigma-\sigma_0)\simeq 1-e^{-k_1(v-v_0)},
\end{equation}
where the subscript $0$ refers to initial values. Since $k_1$ must be positive, when the affine parameter $\sigma$ reaches the finite value
\begin{equation}
\sigma_{\rm h}=\sigma_0+1/(k_1\dot{v}_0),
\end{equation}
the geodesic has reached the limit $v\to\infty$ and, in the absence of singularities, can be extended past this point. Of course, this is the case only if the geodesic stays close enough to $R_{\rm h}(v)$ so as to keep the approximation \eqref{7} valid.\par
As mentioned earlier, if $k_1\neq 0$ then for light rays to be trapped below $r_{\rm h}$ the functions $\delta(v)$ and $d_{\rm h}(v)$ must both tend to zero at least as quickly as an exponential. For example, if
\begin{equation}
\delta(v)=e^{-\alpha v},\qquad d_{\rm h}(v)=e^{-\beta v},
\end{equation}
with $\alpha$ and $\beta$ some positive constants, then the solution for the trajectories of outgoing null geodesics is asymptotically
\begin{equation}\label{13}
r(v)\simeq r_{\rm h}-\frac{1}{\alpha-k_1}e^{-\alpha v}-\frac{k_1r_{\rm h}}{\beta-k_1}e^{-\beta v}+ce^{-k_1 v},
\end{equation}
where $c$ is an integration constant. There are trapped null solutions if $\min(\alpha,\beta)>k_1$. They correspond to the values $c\le0$ for the integration constant ($c=0$ for the horizon itself). The approximation resulting in eq. \eqref{10} is valid for these trajectories (since they approach $R_{\rm h}$), and they are therefore extendable past the $v\to\infty$ limit. At this limit they reach $r=r_{\rm h}$, making this surface a Cauchy horizon, as shown in Fig. \ref{f1}.\par
As for spacelike and timelike geodesics, the equivalent of eq. \eqref{2} is
\begin{equation}\label{14}
\frac{dr}{dv}=F(v,r)\pm\frac{1}{2g(v,r)\dot{v}^2},
\end{equation}
with $+$ for spacelike and $-$ for timelike ones. Since we are interested in the region around $r_{\rm h}$ at large $v$, we only look for geodesics which stick close to this radius asymptotically. Using this as an assumption for the solutions, for $g(v,r)=1$ it is easy to check that with eq. \eqref{10}, $\dot{v}$ diverges quickly enough for the new term in \eqref{14} (with respect to the null case) to become negligible at leading order in the asymptotic expansion. Thus, for every null geodesic of the type \eqref{13} there are also a spacelike and a timelike geodesic with the same approximate expressions. From the signs of the additional term in \eqref{14} it can be seen that further approximation would reveal that in terms of radius the spacelike geodesics are actually slightly above the null ones, while the timelike ones are slightly below. Eq. \eqref{10} is also a valid approximation for the affine parameter of these geodesics, meaning they are also extendable.\par
The same occurs even when $k_1=0$: geodesics which try to escape from the interior region reach the $v\to\infty$ limit in finite affine parameter. For example, if $k_2\neq 0$ we can solve eq. \eqref{9} in the vicinity of $r_{\rm h}$ and see that the value this parameter reaches when $v$ diverges is
\begin{equation}
\sigma_{\rm h}=\sigma_0+\frac{1}{\dot{v}_0k_2(r_{\rm h}-r_0)}.
\end{equation}\par
There are only two exceptions to this scenario of extendable geodesics. The first one is the case in which the function $f$ is constant in $r$, making all coefficients $k_i$ in the expansion \eqref{7} zero; equation \eqref{9} then implies $v\propto\sigma$ and there is no Cauchy horizon for trapped geodesics. The second exception is, in a sense, a generalisation of the first: it is the case in which the function $f$ in non-analytical in the $r$ direction about its minimum, and all its derivatives are zero there. In other words, we can generalise from the case of constant $f$ in $r$ and maintain the non-extendibility by sacrificing the analytic nature of the function. Let us provide an example: suppose we have
\begin{equation}
F(v,r)\simeq e^{-\frac{1}{(r-r_{\rm h})^2}}+\frac{1}{v^n},
\end{equation}
where $r$ and $v$ are expressed in units of some arbitrary length scale. Then the asymptotic solutions for trapped outgoing light rays are
\begin{equation}
r\sim r_{\rm h}-\frac{1}{\sqrt{\log(v-c)}},
\end{equation}
where $c$ is an integration constant and aside from the asymptotic condition $v\gg1$, the range of validity of each solution is $v>c+1$. Along these trajectories eq. \eqref{9} becomes
\begin{equation}
\ddot{v}=\frac{2}{(v-c)[\log(v-c)]^{3/2}}\dot{v}^2,
\end{equation}
the asymptotic solution of which is again $v\propto\sigma$, which we have confirmed both analytically and numerically.\par
In summary, the requirement on $f$ for the geometry to be non-extendable is that all derivatives at its minimum in the direction of decreasing $r$ be zero, either by making the function constant in $r$ or non-analytical. These cases seem rather unphysical, but they do highlight the fact that the presence of a Cauchy horizon depends entirely on the knowledge of the derivatives of $f$ about a single radial point. It is therefore a case in which an arbitrarily small region of the geometry, the description of which may be expected to change in a complete microscopic theory of gravity, affects our picture of the global causal structure of the spacetime.

\subsection{Causal structure with non-vanishing $f$}

If, on the other hand, light rays are trapped not due to a tendency to zero of $f$ but because of increasing values in $g$ at $R_{\rm h}(v)$, tending to a divergence in the limit $v\to\infty$, the situation is quite different. The geodesic equation relating $v$ to the affine parameter $\sigma$ is in this case
\begin{equation}\label{16}
\ddot{v}+\frac{\partial_vg}{g}\dot{v}^2=0.
\end{equation}
For simplicity, we will consider $R_{\rm h}(v)\equiv r_{\rm h}$, since not doing so does not lead to any qualitative changes in the causal structure we will obtain (so long as light-ray trapping is maintained). If $k_1\neq 0$, then we can take $\delta(v)=e^{-\alpha v}$, with $\alpha>k_1$, as it is the slowest allowed approach to zero. The trajectories of trapped outgoing null geodesics are described by \eqref{13} without the $e^{-\beta v}$ term. Then, for large values of $v$ equation \eqref{16} takes the form
\begin{equation}
\ddot{v}=-\frac{2\alpha}{k_1|c|}e^{-(\alpha-k_1)v}\dot{v}^2.
\end{equation}
The solution for the affine parameter $\sigma$ is an exponential integral function with argument proportional to $e^{-(\alpha-k_1)v}$, and at large values of $v$ is approximated by the relation
\begin{equation}\label{18}
\sigma=a_1+a_2v,
\end{equation}
with $a_1$ and $a_2$ being integration constants. Thus in this case the affine parameter of these geodesics reaches infinity at the same time as $v$, meaning that the $r_{\rm h}$ region does not become a Cauchy horizon but a part of future null infinity. If $k_1=0$, there is no change in this behaviour. In fact, eq. \eqref{18} is still the approximate solution relating $v$ to the affine parameter for trapped geodesics at large values of $v$ [e.g. if $k_2\neq0$ and we take again $\delta(v)=e^{-\alpha v}$, the term approximating $v$ on the rhs of eq. \eqref{18} is in this case $\int_1^v\exp(x^2\exp(-\alpha x))\mathop{dx}$].\par

\begin{figure}
	\centering
	\includegraphics[scale=.7]{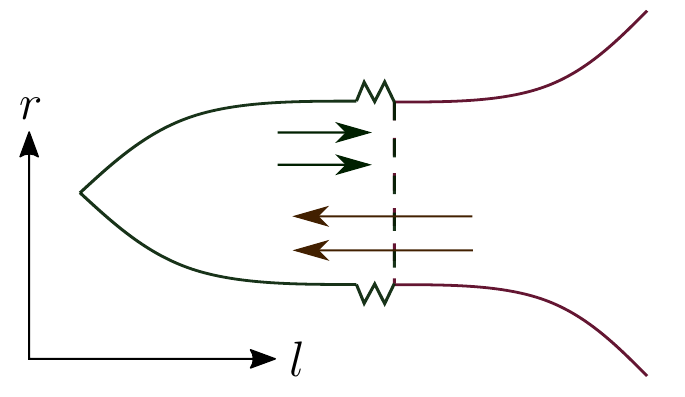}
	\caption{Relation between the radial coordinate $r$ and the proper length in the radial direction $l$ for a geometry in which $g$ tends to a divergence at $r_{\rm h}$. Outgoing light rays become trapped in this infinitely stretching region, while ingoing ones pass right through it.}
	\label{f5}
\end{figure}

This case of diverging $g$ can be interpreted geometrically from these results for null geodesics: space becomes stretched in the radial direction at $r_{\rm h}$ as the proper radial length $l$ approaches a divergence along with $g$. This stretching is sufficiently quick so as to asymptotically freeze these light rays in their approach toward $r_{\rm h}$ (see Fig. \ref{f5}). But the key difference with respect to the previous case is that this occurs without the low values of the redshift function $f$, which results in proper time not being slowed down and observers reaching the asymptotic region $v\to \infty$ in infinite time.\par
As this situation approaches the static case discussed above, one might wonder if ingoing rays would also be affected in a similar manner, becoming unable to cross the $r_{\rm h}$ surface. This turns out not to be the case. The geodesic equation relating the affine parameter to the radial coordinate in $v=\text{const.}$ sections is the same as the second expression in eq. \eqref{5} in terms of $g$, but in this case, close to $r_{\rm h}$ it takes the form
\begin{equation}
\ddot{r}=-\frac{\partial_rg}{g}\dot{r}^2\simeq\frac{m\tilde{k}_m(r-r_{\rm h})^{m-1}}{\tilde{k}_m(r-r_{\rm h})^m+\delta(v)}\dot{r}^2,
\end{equation}
where $m$ is again the order of the first non-zero term in the series expansion of $F$, and $\delta(v)$ is a (small) constant. In contrast to the static case, the rhs is not divergent due to the finite $\delta(v)$ term. Consequently, the affine parameter is finite when crossing $r_{\rm h}$ (e.g. for $m=2$, $r-r_{\rm h}\simeq\sqrt{\delta(v)}\tan[c_1\sqrt{\delta(v)}(\sigma-c_2)]$, with $c_1,c_2$ integration constants; crossing occurs at $\sigma=c_2$).\par
Outgoing light rays become trapped due to the fact that they are moving in a direction in which $g$ increases and $\delta(v)$ decreases, and actually see space stretching as they go. On the other hand, ingoing rays only see a snapshot of a partially stretched geometry, through which they can easily pass given enough time.\par
\begin{figure}
	\centering
	\includegraphics[scale=.57]{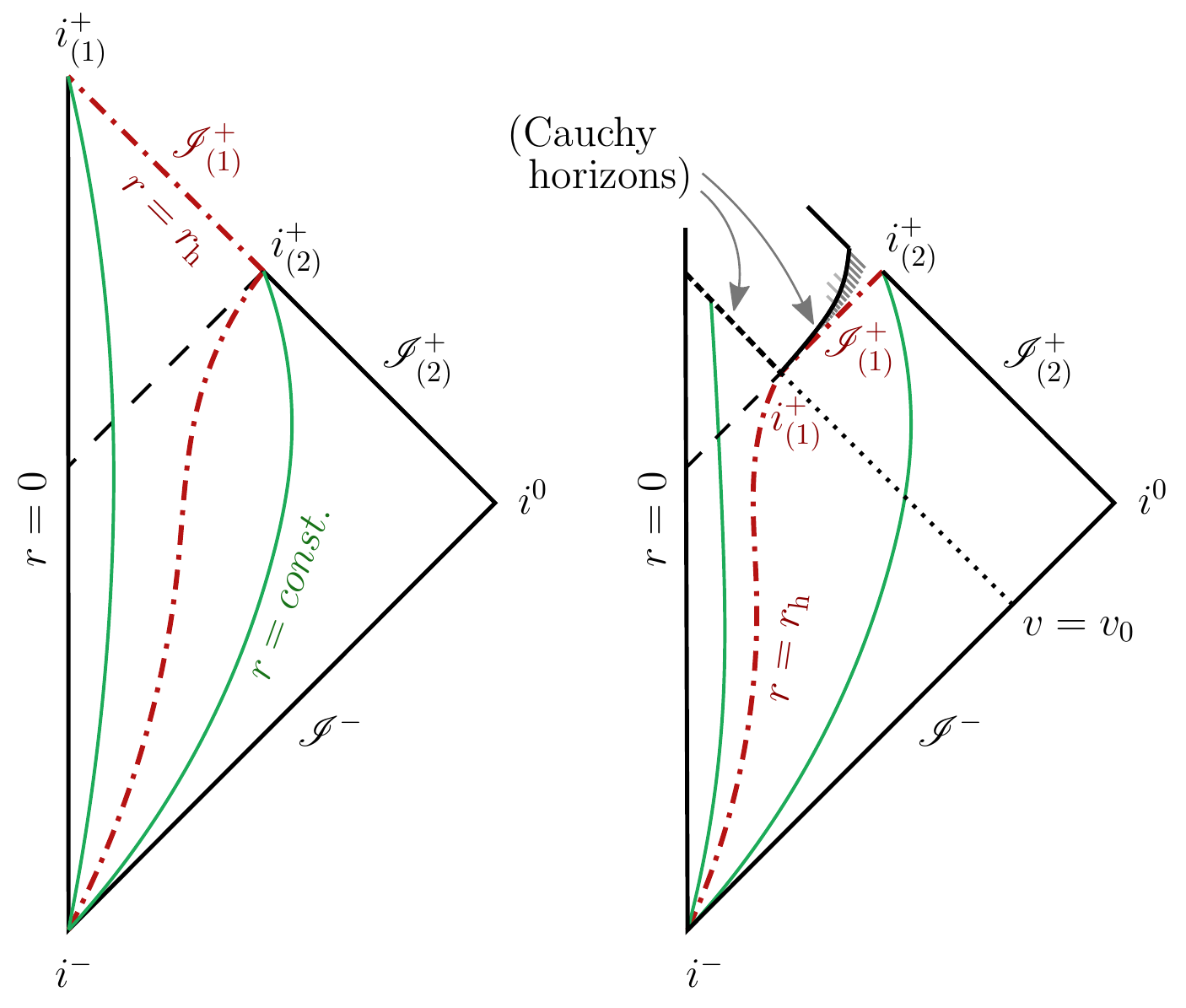}
	\caption{Conformal diagram of the spacetime in which $g$ tends to a divergence. Left: the divergence is reached in infinite time at the surface $r=r_{\rm h}$ (the dash-dotted line), which becomes a separate part of future null infinity for outgoing light rays. This interior null infinity is denoted by $\mathscr{I}_{(1)}^+$, while the exterior one (for escaped light rays) is $\mathscr{I}_{(2)}^+$. Right: the divergence at $r=r_{\rm h}$ is reached at a finite moment $v=v_0$ (and remains thereafter), making $r=r_{\rm h}$ a future null infinity $\mathscr{I}_{(1)}^+$ for ingoing light rays with $v\geq v_0$. The symbols $i^+_{(1)}$ and $i^+_{(2)}$ indicate future timelike infinities for two different sets of observers (for the diagram on the right this is conditional; see discussion below). In general, this latter geometry is extendable past the surface $v=v_0$ for $r<r_{\rm h}$, marked as a Cauchy horizon. The diagram includes what the extension may look like, indicating that to be fixed it requires initial data from another surface, which is effectively another Cauchy horizon in the past of the region.}
	\label{f3}
\end{figure}

The two future null infinities in this spacetime are separated by an event horizon, as shown in Fig. \ref{f3} (left). The exterior one we assume is in an asymptotically flat region at $r\to\infty$, while the interior one (at $r\to r_{\rm h}^-$) has a matter content all the way through, which we will briefly analyse in the next section.\par

\subsection{Diverging $g$ in finite time}

In Fig. \ref{f3}, the diagram on the right represents a case in which the point in time at which $g$ diverges is brought down to a finite value $v=v_0$ (i.e. $\delta(v)=0$ for $v\ge v_0$). The spacetime in this case becomes a combination of the static and asymptotically formed cases, and can help shed light on both.\par
The infinite tube from the static case is now formed dynamically, i.e. space stretches in the radial direction and breaks into two at a point $(v_0,r_{\rm h})$. For $v> v_0$ and $r>r_{\rm h}$ the spacetime is part of the exterior region of the static case, in which the surface $r=r_{\rm h}$ is a future asymptotic region for geodesics which approach it. As for the interior $r<r_{\rm h}$ region, from the moment $v=v_0$ on the evolution is no longer determined by any initial conditions set at any past spacelike 3-surface, making the surface $v=v_0$ for $r<r_{\rm h}$ a Cauchy horizon. The conditions needed to fix a particular extension will generally be determined at a surface which can be thought of as a second Cauchy horizon in the past of the extended region, as shown in Fig. \ref{f3}.\par
The only thing left to analyse in order to complete our picture of this geometry is the point $(v_0,r_{\rm h})$. The first thing to note is that there is no curvature singularity there. Considering that it is the point after which $r=r_{\rm h}$ becomes an asymptotic region, one might initially think of it as part of this region. Then all geodesics which approach it would have an affine parameter which tends to infinity there. We can easily check if this is the case with the geodesic equations.\par
As it turns out, the answer is not that straightforward. Whether this point is part of an asymptotic region for geodesics or not actually depends on how the divergence in $g$ is approached, i.e. how quickly space is stretched. This is encoded in how $\delta(v)$ reaches zero. The details of this calculation are deferred to appendix \ref{ap1}, but the summary is the following: if we have
\begin{equation}
\delta(v)\propto(v_0-v)^n,
\end{equation}
then for all geodesics to have their affine parameters diverge when they reach this point, the inequality
\begin{equation}
n-1\geq\frac{1}{m-1}
\end{equation}
must be satisfied, where $m$ is the lesser of the two numbers corresponding to the orders of the first non-zero coefficients $k_i$ and $\tilde{k}_i$ in the expansion of $F$. If this inequality is not satisfied, then some geodesics will reach this point in finite proper time and will be extendable beyond it. For timelike geodesics, the extensions will be into the interior region beyond the Cauchy horizon.\par
On the other hand, in the case in which the divergence in $g$ is reached in infinite time, i.e. the limit $v_0\to\infty$, all geodesics which approach this point have their affine parameter reach an infinite value at the same rate as $v$ [eq. \eqref{18}]. To understand how the transition from the right to the left diagram in Fig. \ref{f3} occurs, we can think of the fact that in this limit all ingoing light rays make it through $r_{\rm h}$, are reflected at the origin and then become trapped in an approach toward what is essentially the point $(v_0,r_{\rm h})$ from the inside. The point is then stretched to become a future null infinity region for all these rays, as well as a future timelike infinity for geodesics which may approach it from below [$i^+_{(1)}$] or above [$i^+_{(2)}$], as seen in the left diagram in Fig. \ref{f3}. These timelike infinities are also the ones for geodesics which approach the asymptotic region with radii $r<r_{\rm h}$ and $r>r_{\rm h}$ respectively.\par

\section{Energy conditions}\label{Sec:energyconditions}

The physical picture behind the $f\to 0$ type spacetimes is roughly that of a collapse of matter which grinds to a halt asymptotically in time just above its gravitational radius. On the other hand, the $g\to\infty$ type seems to describe a stretching of space in the radial direction in a manner similar to cosmological expansion. These are unusual situations, to say the least, so it is interesting to see whether some of them can be associated with the dynamics of classically reasonable matter, i.e. whether their stress-energy tensors can satisfy any of the energy positivity conditions.\par
We will suppose $w^\mu$ is any timelike or null vector, and without loss of generality we will suppose its angular component is in the $\theta$ direction, resulting in the inequality
\begin{equation}\label{19}
-f(w^v)^2+2gw^vw^r+r^2(w^\theta)^2\le 0.
\end{equation}
The only region in which we have needed to fix the spacetime geometry so far is around $R_{\rm h}(v)$, so we will analyse how matter behaves there, using the expansions \eqref{7} and \eqref{8}.\par
Let us again start with the case $g\simeq 1$. To test the weak (and null) energy condition, we contract the Einstein tensor of \eqref{1} twice with $w^\mu$ and see whether the resulting scalar is positive for all $w^\mu$ satisfying \eqref{19}. Since we will only analyse this condition in the region where we have fixed the geometry, i.e. around the minimum in $F$, we can omit some terms which will not give leading-order contributions and write
\begin{equation}\label{20}
\begin{split}
G_{\mu\nu}w^\mu w^\nu&\simeq\frac{1-r\partial_r f}{r^2}\left[f(w^v)^2-2w^vw^r\right]\\&-\frac{\partial_v f}{r}(w^v)^2+\left(r\partial_r f+\frac{r^2}{2}\partial_r^2f\right)(w^\theta)^2.
\end{split}
\end{equation}
This quantity can be shown to be positive in many cases with some simplifications (we will not attempt to derive the most general conditions for positivity). Particularly, let us assume that $\partial_v f$ is negative [for which $\delta(v)$ must decrease more slowly than $d_{\rm h}(v)$]. Then the term with this partial derivative will be positive and can safely be ignored. With the inequality \eqref{19}, the sufficient conditions for the rest of the terms on the rhs of \eqref{20} to be positive turn out to be
\begin{equation}\label{wec}
r_{\rm h}\partial_r f< 1\quad\text{and}\quad \frac{r_{\rm h}^2}{2}\partial_r^2f> -1,
\end{equation}
which can be satisfied or violated with an appropriate choice of coefficients in \eqref{7} and \eqref{8}. If they are satisfied, then any timelike observer around this region of ``slowed down gravitational collapse" will see a matter distribution with positive energy density.\par
With a similar analysis, it can be shown that the strong energy condition ($R_{\mu\nu}w^\mu w^\nu\ge 0$) can be satisfied around $r_{\rm h}(v)$ if
\begin{equation}\label{sec}
\partial_v f<0\quad\text{and}\quad\frac{r_{\rm h}^2}{2}\partial_r^2f\ge \max(-r_{\rm h}\partial_rf,-1).
\end{equation}
As for the dominant energy condition (that is, requiring that the momentum flux $-T^\mu_{\hphantom{\mu}\nu}w^\nu$ be causal and future-pointing), it can be satisfied if the weak energy condition is, and
\begin{equation}
(1-r_{\rm h}\partial_rf)^2\ge\left(r_{\rm h}\partial_rf+\frac{r_{\rm h}^2}{2}\partial_r^2f\right)^2,
\end{equation}
which can again be achieved with an appropriate choice of coefficients in \eqref{7} and \eqref{8}.\par
This result implies that it is not necessary to violate energy conditions locally in order to generate the $f\to0$ type geometry, but it does not guarantee that for the whole of our spacetime construction. Indeed, in most cases the interior of this geometry ($r<r_{\rm h}$) resembles that of a regular black hole, where some energy conditions are usually broken around the origin $r=0$~\cite{Ansoldi,Bardeen,BeatoGarcia,Hayward}.\par
In the local region where energy conditions can be satisfied, it may also be interesting to see what form the energy density and pressure perceived by an observer can take. Let us consider an observer freely falling in the radial direction, who has a four-velocity $w^\mu$ and, for simplicity, at the moment of crossing $r_{\rm h}$ is moving in the $v$-direction with $w^v\simeq 1$. Then, taking again $g\simeq 1$, the energy density seen by this observer when approaching from the outside is approximately
\begin{equation}
\rho\simeq\frac{1}{8\pi G}\left(\frac{1}{r_{\rm h}^2}-\frac{\partial_rf}{r_{\rm h}}\right)
\end{equation}
Note that the condition for $\rho$ to be positive coincides with the first condition in \eqref{wec} (the second condition there is necessary for tangentially moving observers). The radial pressure seen by this observer is
\begin{equation}
p_{\rm r}\simeq -\rho,
\end{equation}
and the tangential pressure is
\begin{equation}
p_\theta\simeq \frac{1}{8\pi G r_{\rm h}^2}\left(r_{\rm h}\partial_rf+\frac{r_{\rm h}^2}{2}\partial_r^2f\right).
\end{equation}
One may note that this expression is positive if the second inequality in \eqref{sec} is satisfied along with the one for $\rho>0$. From all this we see that, although the energy conditions can be satisfied, the corresponding matter content is classically rather strange: it has a pressure which is generally anisotropic, and in some cases the sign of its radial component is opposite to that of its tangential component.\par
However, if our question is whether these geometries are physically reasonable, even locally, this analysis is incomplete. Due to the causal structure involved, additional considerations must be taken into account. On the one hand, it is well-known that the presence of a Cauchy horizon in a solution of the Einstein equations generally indicates that this solution is unstable under perturbations \cite{PoissonIsrael89}. And even if we ignore the possibility of classical perturbations, if we define a quantum field on this spacetime, an analysis based on semiclassical gravity reveals an even greater instability around Cauchy horizons~\cite{BalbinotPoisson93}. On the other hand, in \cite{BBCG19} we showed that in most of these spacetimes there also seem to be large semiclassical corrections due to the formation of the event horizon itself. Therefore, a geometry of this type satisfying reasonable energy conditions may not be a self-consistent solution of the semiclassical Einstein equations. Conversely, if the stress-energy needed to generate it were not sensible on a purely classical level, it would not be enough to discard it as a solution in semiclassical gravity, which is known to have no regard for classical energy conditions~\cite{Visser2}. A complete analysis of the self-consistency of this type of geometry is, however, beyond the scope of this work.\par
As for the geometries in which $g$ tends to a divergence, it turns out that they generally violate even the weak energy condition. To see this we can write down
\begin{equation}
\begin{split}
G_{\mu\nu}w^\mu w^\nu&=\left(\frac{1}{r^2}+\cdots\right)\left[f(w^v)^2-2gw^vw^r\right]\\&\quad+\left(\frac{\partial_rg\partial_vg}{g^3}+\cdots\right)r^2(w^\theta)^2+\\&\quad+\frac{2f\partial_vg}{rg^2}(w^v)^2+\frac{2\partial_rg}{rg}(w^r)^2.
\end{split}
\end{equation}
The first three terms can be made positive for all $w^\mu$ with a particular choice of $g$, but with the last term it is no longer possible. Particularly, if the last term is negative (which it is for $r>r_{\rm h}$), then any attempt to compensate it with the other terms fails for some choice of vector $w^\mu$. From a more physical perspective, this implies that observers moving sufficiently fast in the radial direction (which becomes increasingly difficult as space stretches, i.e. it requires them to approach the speed of light) may see a negative energy density content. Thus it appears these spacetimes are not ones we may expect to form from the dynamics of exclusively classical matter.

\section{Hawking temperature}\label{s5}

If we define a quantum field on top of these spacetimes, the magnitude of the quantum contribution to the stress-energy content depends greatly on the presence of Hawking radiation and the value of its temperature, as discussed in \cite{BBGJ16,BBCG19}. In this work we have used a slightly more general family of geometries, so we will present a general method for calculating the asymptotic effective temperature function (ETF) of the Hawking radiation generated by these geometries which requires only the approximate asymptotic solutions for the trajectories of outgoing null geodesics in a neighbourhood of their event horizons.\par
The ETF was introduced in \cite{Barceloetal2011a} and is given by
\begin{equation}\label{25}
\kappa_{\rm in}^{\rm out}\equiv-\left.\frac{d^2u_{\rm in}}{du_{\rm out}^2}\right/\frac{du_{\rm in}}{du_{\rm out}},
\end{equation}
where the ``in" and ``out" indices refer to the asymptotically flat regions at past and future null infinities: the coordinates are proportional to the natural Minkowskian coordinates at these regions, and the indices of $\kappa$ refer to the difference between the two natural Minkowskian vacuum states (particularly, how the ``in" region vacuum state is seen as a flux of particles when it evolves and reaches the ``out" region). If this function is approximately constant for a long enough period of time~\cite{Barceloetal2011a}, then during this period the geometry will create particles with a Planckian spectrum with temperature $\kappa_{\rm in}^{\rm out}/2\pi$ in natural units.\par
This function depends only on the quotient $f/g$, so the calculation is the same for the two types of geometries we have considered. We will assume $k_1$ and $\tilde{k}_1$ are non-zero, as the case in which either one is zero can be obtained as a limit from the final result. We will also assume that $\delta(v)$ and $d_{\rm h}(v)$ both decrease as exponentials, since it is the slowest allowed approach to zero for light-ray-trapping to occur in this case, and also because the case of a faster approach can again be obtained from the same result.\par
To calculate the ETF, we need to obtain the trajectories of outgoing null geodesics in a small region around the spatial minimum of $F$. For this we can make use of the solution \eqref{13} for $r<R_{\rm h}$ and its analogue with $k_1\to -\tilde{k}_1$ for $r>R_{\rm h}$. We will take the small region $(r_{\rm h}-\epsilon,r_{\rm h}+\tilde{\epsilon})$, with $\epsilon$ and $\tilde{\epsilon}$ arbitrarily small positive constants (with the condition that time has advanced enough for $R_{\rm h}$ to be inside this radial interval). We call $v_\epsilon$ the time at which a particular ray crosses $r_{\rm h}-\epsilon$, $v_{\rm h}$ the time when it crosses $R_{\rm h}$, and $v_{\tilde{\epsilon}}$ the instant it crosses $r_{\rm h}+\tilde{\epsilon}$. For our purposes, the labels $v_\epsilon$ and $v_{\tilde{\epsilon}}$ represent the $u_{\rm in}$ and $u_{\rm out}$ ones respectively.\par
From the solutions \eqref{13}, a straightforward calculation leads to the asymptotic (in $v$) result
\begin{equation}\label{26}
\frac{dv_\epsilon}{dv_{\rm h}}\sim \beta r_{\rm h}e^{-(\beta-k_1)v_{\rm h}}+e^{-(\alpha-k_1)v_{\rm h}},
\end{equation}
where we have omitted a proportionality constant. From here on we must decide which of these two exponentials dominates at large time, i.e. which one decays slower. If $\alpha<\beta$, then the first one dominates, and we also obtain from the exterior solutions the asymptotic relation
\begin{equation}\label{27}
\frac{dv_{\rm h}}{dv_{\tilde{\epsilon}}}\sim \frac{\tilde{k}_1}{\alpha+\tilde{k}_1}.
\end{equation}
On the other hand, if $\beta<\alpha$, then the second exponential in \eqref{26} dominates and the result is the same as \eqref{27}, only substituting $\alpha$ for $\beta$. Defining $\gamma=\min(\alpha,\beta)$ we can proceed with integrating \eqref{27} in generic terms. Doing so and substituting into \eqref{26} we obtain the asymptotic relation between the labels
\begin{equation}
\frac{dv_\epsilon}{dv_{\tilde{\epsilon}}}\sim e^{-\tilde{k}_1\frac{\gamma-k_1}{\gamma+\tilde{k}_1}v_{\tilde{\epsilon}}}.
\end{equation}
The ETF is simply minus the coefficient multiplying $v_{\tilde{\epsilon}}$ in the exponential,
\begin{equation}\label{29}
\kappa_{\rm in}^{\rm out}\sim \tilde{k}_1\frac{\gamma-k_1}{\gamma+\tilde{k}_1}.
\end{equation}
If either $\delta(v)$ or $d_{\rm h}(v)$ decays quicker than an exponential, then the limit $\alpha\to\infty$ or $\beta\to\infty$ can be taken, respectively. Eq. \eqref{29} still applies if both $\alpha$ and $\beta$ are taken to $\infty$, i.e. if $\gamma\to\infty$, giving simply $\tilde{k}_1$ for the ETF. If the slope on either side of the minimum of $F$ is zero, then the corresponding limits $k_1\to 0$ and $\tilde{k}_1\to 0$ can also be taken, the latter resulting in a zero ETF.\par
One thing which is interesting to note is that the surface gravity of these objects at $r_{\rm h}$ is given by
\begin{equation}
\kappa=\frac{1}{2}\frac{\partial_r f}{g},
\end{equation}
meaning that when $g$ diverges, the surface gravity always tends to zero. More generally, when $g\not\simeq\text{const.}$, there is no longer a direct relation between the surface gravity and the temperature of Hawking radiation corresponding to the horizon. The latter is instead associated with the slope $\partial_r(f/g)$, i.e. the coefficient $k_1$ of the series \eqref{3}.

\section{Conclusions}
The starting point of this paper is a family of geometries which, through their particular asymptotic evolution in time, can behave like black holes and even have event horizons, without ever having formed any trapped surface. Even sharing this characteristic, the family contains various and distinct causal structures and different behaviours in terms of energy conditions and production of Hawking radiation. The family of geometries is divided into two categories.\par
The first one is characterised by its similarity with a spacetime in which a standard black hole is formed, but in our case the formation of its first trapped surface is pushed forward to the future asymptotic region. In other words, the strict formation of a trapped surface is replaced by an appropriately quick tendency to its formation, quick enough that although outgoing radial light rays always have a positive expansion, some move out slowly enough to be trapped inside a finite spatial region until the advanced time $v$ reaches infinity.\par
Analysing the causal structure of this first category of geometries, we find that aside from an event horizon (described by the first trapped outgoing light ray) in almost all cases there is also a Cauchy horizon, beyond which the trapped geodesics are extendable, giving the same causal structure in the future as an extremal charged black hole (though in our case it can be singularity-free). However, we find that there are two exceptions to this scenario in which geodesics are not extendable and there is no Cauchy horizon: the first is a very unique case in which the redshift function $f$ has no variation in the radial direction in sections of constant $v$, while the second (a generalisation of the first) just requires that all derivatives in the inward radial direction from the minimum of $f$ be zero. This latter case involves geometries in which $f$ is not constant in $r$ but is non-analytical. We thus point out the interesting fact that the presence of a Cauchy horizon is deduced form the shape of the geometry in an arbitrarily small region about the minimum of $f$, but has consequences on the global causal structure.\par
The second category of geometries in which outgoing light rays are trapped has a very different physical picture behind it. Instead of a decreasing redshift function $f$, what results in the slow-down of the radial escape of the light rays is an actual stretching of space in the radial direction. The proper length becomes vastly greater that the radial length, tending to a divergence in their relation. One can think of it as an attempt at opening a wormhole with an infinitely long neck. To simulate the asymptotic formation of a trapped surface, this divergence only needs to be reached asymptotically as well. Meanwhile, because the stretching increases in the $v$ direction, ingoing geodesics can enter the trapped region after traversing a long, but finite tube-like structure. The difference with the first category of spacetimes is most clearly manifest in the causal structure: outgoing geodesics which are trapped below some finite radius are now \textit{not} extendable beyond the $v\to\infty$ border, i.e. their affine parameter also reaches infinity.\par
This separation into two categories can be seen as due to the fact that requiring for outgoing null trajectories to be trapped defines only what we call the \textit{generalised redshift function} $F(v,r)$, which amounts to just one of the two degrees of freedom of spherically symmetric geometries. However, the geodesic equations, from which we deduce the causal structure, see both of these degrees of freedom. Thus, different ways of imposing the same behaviour in $F$ result in different behaviours of the geodesic affine parameter.\par
Having studied the causal structure of these spacetimes, we then looked at the matter content which they require as a source in order to be considered solutions of the Einstein equations. The geometries of the first category can be sustained by a matter content which satisfies any of the energy positivity conditions, that is, at least locally around the point of asymptotic horizon formation, where the geometry is specified. On the other hand, the cases of the second category appear to violate even the weak energy condition. They would thus lose their physical significance in a purely classical theory, but we remind the reader that the grounds for this study are originally the analysis of semiclassical effects in geometries with appropriate null geodesic peeling for non-local quantum effects to manifest. The quantum contributions to the stress-energy content are known to violate all energy conditions as well, which calls for a broadening of our physical criteria.\par
Finally, we briefly delved into the quantum effects induced by these geometries. We restricted ourselves to the study of the Hawking radiation they produce, as its relation to the quantum stress-energy tensor was discussed in our previous work \cite{BBCG19}. Within our family of geometries there are interesting examples in which the surface gravity at the horizon is absolutely distinct from the temperature of Hawking radiation at infinity. This happens in general whenever the function $g$ is not constant at the event horizon. The set of geometries analysed in which $g$ diverges at the horizon provides a clear example of the possibility of having Hawking-like radiation even with zero surface gravity. The peeling of geodesics required for having Hawking radiation is provided in this case by the stretching of space itself.

\section{Acknowledgements}
Financial support was provided by the Spanish Government through the projects FIS2017-86497-C2-1-P, FIS2017-86497-C2-2-P (with FEDER contribution), FIS2016-78859-P (AEI/FEDER,UE), and by the Junta de Andalucía through the project FQM219. VB is funded by the Spanish Government fellowship FPU17/04471. RCR acknowledges support from the Preeminent Postdoctoral Program (P$^3$) at UCF. CB acknowledges financial support from the State Agency for Research of the Spanish MCIU through the ``Center of Excellence Severo Ochoa'' award to the Instituto de Astrof\'{\i}sica de Andaluc\'{\i}a (SEV-2017-0709).

\appendix

\section{Geodesics approaching the point of divergent $g$}\label{ap1}
If $g(v,r)$ diverges at a point $(v_0,r_h)$, close to this point we can quite generally assume it has the form
\begin{equation}\label{a1}
g=\frac{1}{a(v_0-v)^n+\tilde{k}_m(r-r_h)^m},
\end{equation}
for which we have also assumed that we are approaching from a smaller $v$ and a larger $r$, with $a$ and $\tilde{k}_m$ being positive constants. What we want to find out is, depending on the values of $n$ and $m$, whether there are geodesics which approach this divergent point, and if there are, whether they take a finite of infinite proper time to reach it.\par
The easiest way to obtain an answer is to assume we already have it, and then check if it is true. In other words, let us first assume that there are timelike geodesics which reach this point at a finite affine parameter $\sigma_0$ as
\begin{align}\label{a2}
v-v_0&=-\beta(\sigma_0-\sigma)^q+\cdots,\\ \label{a3}
r-r_{\rm h}&=\alpha(\sigma_0-\sigma)^p+\cdots,
\end{align}
with $\beta$, $\alpha$, $p$ and $q$ positive constants. The geodesic equations these trajectories must satisfy are
\begin{equation}\label{a4}
\ddot{v}=-\frac{\partial_vg}{g}\dot{v}^2,\qquad \ddot{r}=-\frac{\partial_rg}{g}\dot{r}^2+\frac{f}{g}\ddot{v},
\end{equation}
where we have assumed $f$ is constant. Plugging the expressions \eqref{a2} and \eqref{a3} into these equations, we get the following results for the leading order
\begin{align}\label{a6}
q&=\frac{2}{1-n+n/m},\qquad \\ \label{a7}
p=\frac{n}{m}q&=\frac{2}{1-m+m/n},\\ \label{a8}
\frac{\alpha^m}{\beta^n}&=\frac{a}{\tilde{k}_m}\frac{mn-m+n}{mn+m-n}.
\end{align}
There is a single degree of freedom left in the proportionality coefficients, meaning we have found a whole uniparametric family of solutions. An important point is that these solutions are valid representations of geodesics which reach the point of divergent $g$ only if $q$ and $p$ are positive, which implies the restriction
\begin{equation}\label{a9}
n-1<\frac{1}{m-1}
\end{equation}
for the geometry. The smaller the exponents $n$ and $m$, the quicker the divergence is approached, so this inequality can be interpreted as the fact that geodesics only take a finite time to reach the point if the divergence is generated suddenly enough.\par
On the other hand, if we assume the geodesics take an infinite time to reach the point, say as
\begin{align}\label{a10}
v-v_0&=-\frac{\beta}{\sigma^q}+\cdots,\\ \label{a11}
r-r_{\rm h}&=\frac{\alpha}{\sigma^p}+\cdots,
\end{align}
then the opposite inequality,
\begin{equation}
n-1>\frac{1}{m-1},
\end{equation}
must be satisfied, i.e. the divergence of $g$ must be reached slowly enough. Equations \eqref{a6} and \eqref{a7} now hold with a change of sign of the rhs, and eq. \eqref{a8} holds as such.\par
We may then ask whether such geodesics exist for a geometry which precisely satisfies
\begin{equation}
n-1=\frac{1}{m-1}.
\end{equation}
They do, and they take the form
\begin{align}
v-v_0&=-\beta\mathop{e^{-q\sigma}}+\cdots,\\
r-r_{\rm h}&=\alpha\mathop{e^{-p\sigma}}+\cdots,
\end{align}
i.e. they also take infinite proper time to reach the point but they have a different approach. In this case the restrictions on the coefficients imposed by the geodesic equations are
\begin{equation}
\frac{p}{q}=n-1,\qquad \frac{\alpha^m}{\beta^n}=\frac{a}{\tilde{k}_m}(n-1).
\end{equation}\par
So far we have only considered timelike geodesics which fall into $(v_0,r_{\rm h})$ from larger radii. If we also consider ones which may approach this point from the inside, we obtain some additional solutions. Assuming the point is reached in finite proper time, i.e. taking eqs. \eqref{a2} and \eqref{a3}, the latter with a change of sign for the approach from the inside, we get on the one hand solutions which again satisfy eqs. \eqref{a6}, \eqref{a7} and \eqref{a8} (with $\tilde{k}_m\to k_m$, as we are now on the inside), and on the other we obtain some independent additional solutions which satisfy
\begin{align}\label{a18}
p=\frac{1+n}{1-n},&\qquad q=\frac{1}{1-n},\\ \label{a20}
\frac{\alpha}{\beta^{n-1}}&=\frac{f}{2}\frac{a}{1+n}.
\end{align}
The restriction on the geometry for these solutions to exist is simply
\begin{equation}
n<1.
\end{equation}
This kind of additional solutions also exist if we assume an approach in infinite proper time using eqs. \eqref{a10} and \eqref{a11}, the latter again with a change of sign. They satisfy eqs. \eqref{a18} with a change of sign on the rhs, and eq. \eqref{a20} changing the power of $\beta$ from $n-1$ to $n+1$. The geometries on which these solutions exist only need to satisfy
\begin{equation}
n>1.
\end{equation}\par
The conclusion is that if the geometry is given by \eqref{a1} and satisfies
\begin{equation}
n-1\ge\frac{1}{m-1},
\end{equation}
then all geodesics which approach $(v_0,r_{\rm h})$ have their affine parameter tending to infinity. If the opposite relation is satisfied, but $n>1$, then depending on their approach some geodesics will reach this point in finite affine parameter, and some others in infinite. We also remind the reader that throughout the main text we assumed $m\ge1$, which is required for light-ray trapping if the approach toward the divergence in $g$ occurs in infinite advanced time $v$. If we want to relax this restriction in the finite-time diverging case, then the solutions obtained at the beginning of this appendix for an approach in finite proper time \eqref{a2}, \eqref{a3} only exist if the additional restriction $m>1-1/(n+1)$ is satisfied. Also, the ingoing null geodesic which reaches this point does so in finite affine parameter if $m<1$, whereas it always did so in infinite time (just as in the static case) when $m\ge1$.

\nocite{*}
\bibliography{Bibliografia}
\bibliographystyle{ieeetr}

\end{document}